\documentclass[11pt,a4paper]{article}
\usepackage[english]{babel}
\usepackage{amsmath,amsthm,amssymb,epsfig,latexsym}
\usepackage{color}
\usepackage{ulem}
\usepackage[numbers,square,sort&compress]{natbib}

\newcommand{\so}{\scriptscriptstyle \rm I}
\newcommand{\st}{\scriptscriptstyle \rm I\hspace{-1pt}I}
\newcommand{\sth}{\scriptscriptstyle \rm I\hspace{-1pt}I\hspace{-1pt}I}
\newcommand{\qo}{\rm i}
\newcommand{\qt}{\rm ii}
\newcommand{\qth}{\rm iii}
\newcommand{\be}[1]{\begin{equation}\label{#1}}
\newcommand{\ba}[1]{\begin{multline}\label{#1}}
\newcommand{\ee}{\end{equation}}
\newcommand{\ea}{\end{eqnarray}}

\newcommand{\tr}{\mathop{\rm tr}}

\newcommand{\Tg}{\widetilde{T}}
\newcommand{\Ag}{\widetilde{A}}
\newcommand{\Bg}{\widetilde{B}}
\newcommand{\Cg}{\widetilde{C}}
\newcommand{\Dg}{\widetilde{D}}

\newcommand{\kk}{\kappa}

\newcommand{\bu}{\bar u}

\newcommand{\bw}{\bar w}

\newtheorem{lemma}{Lemma}[section]
\newtheorem{prop}{Proposition}

\newtheorem{thm}{Theorem}[section]

\begin{document}


\vspace{12pt}

\begin{center}
\begin{LARGE}
{\bf A note on $\mathfrak{gl}_2$-invariant Bethe vectors}
\end{LARGE}

\vspace{40pt}

\begin{large}
{S. Belliard${}^{\#\,\dagger}$ and N.~A.~Slavnov${}^\ddagger$\  \footnote{samuel.belliard@gmail.com, nslavnov@mi.ras.ru}}
\end{large}

 \vspace{12mm}

${}^\#$ {\it Sorbonne Universit\'e, CNRS, Laboratoire de Physique Th\'eorique et Hautes Energies,  LPTHE,  F-75005, Paris, France}

  \vspace{4mm}

${}^\dagger$ {\it Institut de Physique Th\'eorique, DSM, CEA, URA2306 CNRS Saclay, F-91191, Gif-sur-Yvette, France}

 \vspace{4mm}

${}^\ddagger$  {\it Steklov Mathematical Institute of Russian Academy of Sciences, Moscow, Russia}

\end{center}

\vspace{4mm}


\begin{abstract}
We consider $\mathfrak{gl}_2$-invariant quantum integrable models solvable by the algebraic Bethe ansatz.
We show that the form of on-shell Bethe vectors is preserved under certain twist transformations of the monodromy
matrix. We also derive the actions of the twisted monodromy matrix entries onto twisted off-shell Bethe vectors.
\end{abstract}

\vspace{5mm}

\section{Introduction}

Recently, a new method for constructing Bethe vectors in quantum $\mathfrak{gl}_N$-invariant spin chains  was proposed
in \cite{GroLMS17}. The main observation of this work is that an operator that is used to build a basis in
the Separation of Variables (SoV) approach can also be used within the framework of the Algebraic Bethe Ansatz (ABA) to construct
a basis of the transfer matrix eigenvectors.

To illustrate this statement we consider a $\mathfrak{gl}_2$-invariant spin chain with a monodromy matrix
\be{T}
T(z)=\begin{pmatrix} A(z)& B(z)\\ C(z)& D(z)
\end{pmatrix}.
\ee
Within the framework of ABA \cite{FadST79,BogIK93L,FadLH96}, the eigenstates of the corresponding quantum Hamiltonian can be obtained by the successive
action of the $B$ operator on a referent state $|0\rangle$
\be{EigStat}
B(u_1)\dots B(u_n)|0\rangle,
\ee
provided the parameters $\{u_1,\dots,u_n\}$ satisfy a system of Bethe equations (see \eqref{BE1} below).

On the other hand, to consider the spectrum problem within the framework of SoV approach \cite{Skl90,Skl91,Skl95}, one should make a twist transformation of the monodromy
matrix \eqref{T}:
\be{Tg0}
\kappa T(z)\kappa^{-1}=\Tg(z)=\begin{pmatrix} \Ag(z)& \Bg(z)\\ \Cg(z)& \Dg(z)
\end{pmatrix},
\ee
where $\kappa$ is an invertible $c$-number matrix. For some specific representation of the monodromy matrix, the SoV basis is associated with the operator-valued roots of equation $\Bg(u)=0$. The twist matrix and the representation are chosen in such a way that $\Bg(u)$ has a simple spectrum \cite{GroLMS17,JKKS17}
necessary for the implementation of the SoV approach.

It was shown in \cite{GroLMS17} that the states \eqref{EigStat} also can be written in terms of the new $\Bg$ operators as
\be{tEigStat}
\Bg(u_1)\dots \Bg(u_n)|0\rangle\propto B(u_1)\dots B(u_n)|0\rangle.
\ee
Here the parameters $\{u_1,\dots,u_n\}$ also should satisfy the system of Bethe equations. In other words, the functional
dependence of the Hamiltonian eigenstates on the $B$ operator is invariant under the twist transformation of the monodromy
matrix.

This fact was proved in \cite{GroLMS17} via the SoV method. In this paper we prove the property \eqref{tEigStat} by means of ABA.

Actually, we give two proofs. The first one is elementary. In fact, it literally  mimics the well known classical scheme of ABA \cite{FadST79}.
The main point is that the referent state is no longer an eigenvector of the diagonal elements of the
twisted monodromy matrix. We show, however, that this fact is not crucial here. On the contrary, the key point is that the twist transformation
\eqref{Tg0} preserves the trace of the monodromy matrix
\be{trans}
 \mathcal{T}(z)=\tr \Tg(z)=\tr T(z).
\ee
Furthermore, we do not use any specific representation of the
algebra of the $T_{ij}$ operators. Thus, we show that \eqref{tEigStat} is valid not only for spin chains, but
for any ABA-solvable model.

The second proof is more complex. For this proof, one should explicitly compute the multiple action of the operators $\Bg$ on the
referent state for generic complex $\{u_1,\dots,u_n\}$. The advantage of this way is that one can explicitly see how the state
$\Bg(u_1)\dots \Bg(u_n)|0\rangle$ turns into the state $B(u_1)\dots B(u_n)|0\rangle$ if the Bethe equations are imposed.

We also found it necessary to give this complex proof, because it has a direct application to the Modified Algebraic Bethe Ansatz
(MABA) \cite{BC13,Bel14,BP151,ABGP15,BP152}. Within the framework of this method one considers more sophisticated twist transformation
$\Tg(z)=\kappa_1 T(z)\kappa_2$ with  $\kappa_2\ne\kappa^{-1}_1$. Generically, this transformation does not preserve the trace of the monodromy
matrix, leading to the break of the $U(1)$ symmetry. As a result, the property \eqref{tEigStat} is no longer true for these models, therefore,
one should find a way to describe the eigenstates of the corresponding quantum Hamiltonians. Our second proof gives a tool for this description.

This paper is organized as follows. In  section~\ref{ABA} we introduce a relevant notation and recall the classical scheme of the ABA. In section~\ref{NeABA} we present the new scheme of the ABA for the monodromy matrix \eqref{Tg0} and give an elementary proof of \eqref{tEigStat}.
Section~\ref{S-SP} is devoted to the second proof. The most complex part of it is moved to appendix~\ref{Proof}.

\section{Algebraic Bethe Ansatz}\label{ABA}

We briefly recall the classical scheme of the ABA (see \cite{FadST79,BogIK93L,FadLH96} for more details).
The main objects of this method are a monodromy matrix $T(u)$, an $R$-matrix, and a vacuum vector $|0\rangle$ (referent state).
In the case under consideration the monodromy matrix is a $2\times 2$ matrix \eqref{T},
whose entries are operators acting in some Hilbert space $\mathcal{H}$. Commutation relations between $T_{ij}$ are given
by an $RTT$-relation
\be{RTT}
R(u,v)\bigl(T(u)\otimes I\bigr)\bigl(I\otimes T(v)\bigr) = \bigl(I\otimes T(v)\bigr)\bigl(T(u)\otimes I \bigr) R(u,v),
\ee
where $R$-matrix $R(u,v)$ is a $4\times 4$ $c$-number matrix satisfying the Yang--Baxter equation.
In particular, we consider
\be{Rmat}
R(u,v)=\mathbf{I}+g(u,v)\mathbf{P}, \qquad g(u,v)=\frac{c}{u-v}.
\ee
Here $\mathbf{I}$ is the identity matrix, $\mathbf{P}$ is the permutation matrix and $c$ is a constant.
It follows immediately from \eqref{RTT} that a transfer matrix $\mathcal{T}(z)=\tr T(z)=A(z)+D(z)$ possesses a property
$[\mathcal{T}(y),\mathcal{T}(z)]=0$ for arbitrary $y$ and $z$, and thus, it can be considered as a generating function
of integrals of motion of a quantum integrable model.

Let $a(z)$ and $d(z)$ be some functions dependent on a concrete model. We assume that  there exists a vacuum vector $|0\rangle\in \mathcal{H}$
such that
\be{vac}
A(z)|0\rangle=a(z)|0\rangle, \qquad D(z)|0\rangle=d(z)|0\rangle, \qquad C(z)|0\rangle=0.
\ee
The ABA allows one to find the eigenvectors of the transfer matrix. These vectors are commonly called
{\it on-shell Bethe vectors}.
Within the framework of this method, the states of the space $\mathcal{H}$ are generated by multiple action of the
operator $B(u)$ onto the  vacuum vector $|0\rangle$ as in \eqref{EigStat}.

Before describing the basic procedure of ABA, we introduce a new notation. First of all, we need one more rational
function
\be{f}
f(u,v)=1+g(u,v)=\frac{u-v+c}{u-v}.
\ee
Below we will consider a set of parameters $\{u_1,\dots,u_n\}$, which we denote by a bar: $\bu=\{u_1,\dots,u_n\}$. We agree upon
that the notation $\bu_k$ refers to a set that is complementary to the element $u_k$, that is, $\bu_k=\bu\setminus u_k$.
We use a shorthand notation for the products over the sets $\bu$ and $\bu_k$:
\be{shn}
B(\bu)=\prod_{j=1}^nB(u_j), \qquad f(z,\bu)=\prod_{j=1}^n f(z,u_j), \qquad f(\bu_k,u_k)=\prod_{\substack{j=1\\j\ne k}}^nf(u_j,u_k),
\ee
and so on. Note that due to commutativity of the $B$-operators the first product in \eqref{shn} is well defined.

Now we are in position to describe the classical result of ABA \cite{FadST79,BogIK93L,FadLH96}. We are looking
the eigenstates of the transfer matrix in the form
\be{Eig}
|\Psi_n(\bu)\rangle=B(\bu)|0\rangle, \qquad n=0,1,\dots.
\ee
If the parameters $\bu$ are generic
complex numbers, then the state \eqref{Eig} is called an {\it off-shell Bethe vector}.
However, if the parameters $\bu$ satisfy a system of Bethe equations
\be{BE1}
a(u_k)f(\bu_k,u_k)=d(u_k)f(u_k,\bu_k), \qquad k=1,\dots, n,
\ee
then the vector $|\Psi_n(\bu)\rangle$ becomes an {\it on-shell Bethe vector}, that is, an eigenvector
of the transfer matrix.

The proof of this statement is based on the commutation relations between the operators $A(z)$, $D(z)$, and
$B(\bu)$. Namely, if the $R$-matrix has the form \eqref{Rmat}, then
\begin{equation}\label{ComADB}
\begin{aligned}
A(z)B(\bu)=B(\bu)A(z)f(\bu,z)
+\sum_{k=1}^n B(z)B(\bu_k)A(u_k)g(z,u_k)f(\bu_k,u_k),\\
D(z)B(\bu)=B(\bu)D(z)f(z,\bu)
+\sum_{k=1}^n B(z)B(\bu_k)D(u_k)g(u_k,z)f(u_k,\bu_k).
\end{aligned}
\end{equation}
We stress that equations \eqref{ComADB} are direct consequences of the $RTT$-relation \eqref{RTT}.

Acting with equations \eqref{ComADB} onto $|0\rangle$ and using \eqref{vac} we obtain
\begin{equation}\label{ActADB}
\mathcal{T}(z)B(\bu)|0\rangle=\Lambda_0B(\bu)|0\rangle
+\sum_{k=1}^n \Lambda_kB(\bu_k)B(z)|0\rangle,
\end{equation}
where
\begin{equation}\label{act-trst}
\begin{aligned}
&\Lambda_0=a(z)f(\bu,z)+d(z)f(z,\bu),\\
&\Lambda_k=g(z,u_k)\Bigl(a(u_k)f(u_j,u_k)-d(u_k)f(u_k,\bu_k)\Bigr),\qquad k=1,\dots,n.
\end{aligned}
\end{equation}
It is clear that a requirement $\Lambda_k=0$ for $k=1,\dots,n$ is equivalent to the system of Bethe equations \eqref{BE1}.
Then it follows from \eqref{ActADB} that the vector $B(\bu)|0\rangle$ is the eigenvector of the transfer matrix $\mathcal{T}(z)$
with the eigenvalue $\Lambda_0$.

\section{Elementary proof} \label{NeABA}

Let $\kappa_1$ and $\kappa_2$ be a $c$-number  $2\times 2$ matrices, such that $[R(u,v),\kappa_i\otimes\kappa_i]=0$, for $i=1,2$. Then, it is well known
(see e.g. \cite{FadST79,BogIK93L,FadLH96}) that a {\it twisted monodromy matrix} $\Tg(u)=\kappa_1 T(u)\kappa_2$   also satisfies the $RTT$-relation \eqref{RTT}.
It is easy to see that in the case of the $R$-matrix \eqref{Rmat} the condition
$[R(u,v),\kappa_i\otimes\kappa_i]=0$ holds for any $\kappa_i\in \mathfrak{gl}_2$. Therefore, the $R$-matrix \eqref{Rmat} is called
$\mathfrak{gl}_2$-invariant $R$-matrix.

Consider a special twist \eqref{Tg0}, where $\kk$ is an invertible matrix. As we have already mentioned,
this twist transformation preserves the transfer matrix $\mathcal{T}(z)$. However,
if the twist matrix $\kappa$ is not diagonal, then the entries of the twisted monodromy matrix \eqref{Tg0}
are linear combinations of the original $A$, $B$, $C$, and $D$ operators. Thus, their actions on the vacuum vector $|0\rangle$ is no longer given
by equations \eqref{vac}.
Nevertheless, if  $\kk_{11}\ne 0$, then the on-shell Bethe vectors  can be presented in terms of the $\Bg$ operators as in \eqref{tEigStat},
provided the parameters $\bu$ satisfy the same system of Bethe equations \eqref{BE1}.

At the first sight equation \eqref{tEigStat} look strange
and even mysterious, as the vector $\Bg(\bu)|0\rangle$ is a linear combination of states of the form \eqref{Eig},
in which the states depend on all possible subsets of the set $\bu$.
However, from the point of view of the ABA, this result
directly follows from the $RTT$-relations \eqref{RTT} and the fact that $\tr \Tg(u)=\tr T(u)$. It is valid for much wider
class of models, but not only for spin chains.

Let us turn back to the twisted monodromy matrix \eqref{Tg0} and consider the action on the vacuum vector of the new diagonal operators $\Ag(z)$ and $\Dg(z)$. Let
\be{K}
\kappa =\begin{pmatrix} \kk_{11}& \kk_{12}\\ \kk_{21}& \kk_{22}
\end{pmatrix},
\ee
where $\kk$ is invertible and $\kk_{11}\ne 0$. Without loss of generality, we
assume that $\det \kappa=1$.
\begin{prop}\label{Vac act} The new operators \eqref{Tg0} act on the vacuum vector  $|0\rangle$ as follows:
\be{Agv1}
\Ag(z)|0\rangle=a(z)|0\rangle-\frac{\kk_{21}}{\kk_{11}}\Bg(z)|0\rangle, \qquad \Dg(z)|0\rangle=d(z)|0\rangle+\frac{\kk_{21}}{\kk_{11}}\Bg(z)|0\rangle.
\ee
\end{prop}
{\sl Proof} It follows from \eqref{Tg0} that
\be{Ag}
\Ag(z)=\kk_{11}\kk_{22}A(z)-\kk_{11}\kk_{21}B(z) +\kk_{12}\kk_{22}C(z) -\kk_{12}\kk_{21}D(z),
\ee
\be{Dg}
\Dg(z)=\kk_{11}\kk_{22}D(z)+\kk_{11}\kk_{21}B(z) -\kk_{12}\kk_{22}C(z) -\kk_{12}\kk_{21}A(z),
\ee
and
\be{Bg}
\Bg(z)=\kk_{11}^2B(z)+\kk_{11}\kk_{12}\bigl(D(z) -A(z)\bigr) -\kk_{12}^2C(z).
\ee
Equation \eqref{Bg} shows the importance of the condition $\kk_{11}\ne 0$. Otherwise, for $\kk_{11}=0$, the new creation operator $\Bg(z)$ would be proportional to the annihilation operator $C(z)$.
Acting with \eqref{Ag}--\eqref{Bg} on the vacuum vector via \eqref{vac}  we after elementary linear algebra arrive at \eqref{Agv1}.
\qed
\vspace{2mm}

We can explicitly see that the vacuum vector $|0\rangle$ remains the eigenvector of the transfer matrix and additional terms in
the actions of the new diagonal operators $\Ag(z)$ and $\Dg(z)$ compensate each other. We will show that the same compensation takes place
in the action of the transfer matrix on the state $\Bg(\bu)|0\rangle$ with arbitrary parameters $\bu$.

Since the twisted monodromy matrix $\Tg(z)$ satisfies the $RTT$-relation \eqref{RTT}, we immediately obtain commutation relations
of the operators $\Ag$ and $\Dg$ with the product of the operators $\Bg$. They are given by equations
\eqref{ComADB}, in which one should replace $\{A,D,B\}$ with $\{\Ag,\Dg,\Bg\}$. Acting with these formulas onto the vacuum vector
we arrive at the following

\begin{prop}\label{OffBV act} The actions of the new  operators $\Ag(z)$ and $\Dg(z)$ on the state $\Bg(\bu)|0\rangle$ are given by
\be{OffBV1}
\begin{aligned}
\Ag(z)\Bg(\bu)|0\rangle&=-\frac{\kk_{21}}{\kk_{11}}\Bg(z)\Bg(\bu)|0\rangle +a(z)f(\bu,z)\Bg(\bu)|0\rangle\\
&\qquad\qquad\qquad+\sum_{k=1}^ng(z,u_k)a(u_k)f(\bu_k,u_k)\Bg(z)\Bg(\bu_k)|0\rangle,
\end{aligned}
\ee
\be{OffBV2}
\begin{aligned}
\Dg(z)\Bg(\bu)|0\rangle&=\frac{\kk_{21}}{\kk_{11}}\Bg(z)\Bg(\bu)|0\rangle+d(z)f(z,\bu)\Bg(\bu)|0\rangle \\
&\qquad\qquad\qquad+\sum_{k=1}^ng(u_k,z)d(u_k)f(u_k,\bu_k)\Bg(z)\Bg(\bu_k)|0\rangle.
\end{aligned}
\ee
\end{prop}
{\sl Proof}. Let us consider the first action. Due to the first equation \eqref{ComADB} we have
\begin{equation}\label{ActtA}
\Ag(z)\Bg(\bu)|0\rangle=f(\bu,z)\Bg(\bu)\Ag(z)|0\rangle+\sum_{k=1}^n g(z,u_k)f(\bu_k,u_k)\Bg(z)\Bg(\bu_k)\Ag(u_k)|0\rangle.
\end{equation}
Now we act with $\Ag(z)$ and $\Ag(u_k)$ onto $|0\rangle$ via \eqref{Agv1}:
\begin{multline}\label{Actt2}
\Ag(z)\Bg(\bu)|0\rangle= a(z)f(\bu,z)\Bg(\bu)|0\rangle+\sum_{k=1}^n a(u_k)g(z,u_k)f(\bu_k,u_k)\Bg(z)\Bg(\bu_k)|0\rangle\\
-\frac{\kk_{21}}{\kk_{11}}\Big(f(\bu,z)+\sum_{k=1}^ng(z,u_k)f(\bu_k,u_k)\Big)\Bg(z)\Bg(\bu)|0\rangle.
\end{multline}
We see that in comparison with the usual action of the operator $A(z)$ onto off-shell Bethe vector $B(\bu)|0\rangle$ we obtain
an additional contribution with $n+1$ operators $\Bg$.  This new term arises due to the new action on the vacuum vector \eqref{Agv1}.
One can easily convince himself that
\be{PDF}
f(\bu,z)-1=\sum_{k=1}^n g(u_k,z)f(\bu_k,u_k),
\ee
because the rhs of \eqref{PDF} is nothing but a partial fraction decomposition of the lhs.
Thus, using \eqref{PDF} we immediately obtain the result.

The action \eqref{OffBV2} can be considered exactly in the  same manner. In this case one should use a partial fraction decomposition
\be{IdA1}
f(z,\bu)-1=\sum_{k=1}^ng(z,u_k)f(u_k,\bu_k).
\ee
Thus, the proof of proposition~\ref{OffBV act} is completed.
\qed
\vspace{2mm}

\begin{thm}\label{P-action}
The action of the transfer matrix $\mathcal{T}(z)$ onto the state $\Bg(\bu)|0\rangle$ reads
\begin{equation}\label{ActtADB}
\mathcal{T}(z)\Bg(\bu)|0\rangle=\Lambda_0\Bg(\bu)|0\rangle
+\sum_{k=1}^n \Lambda_k\Bg(z) \Bg(\bu_k)|0\rangle,
\end{equation}
where $\Lambda_0$ and $\Lambda_k$ are given by \eqref{act-trst}.
\end{thm}
{\sl Proof}. This theorem is a direct consequence of proposition~\ref{OffBV act} and the fact that the twist transformation
\eqref{Tg0} preserves the transfer matrix.
\qed
\vspace{2mm}

Thus, theorem~\ref{P-action} states that the action of the transfer matrix $\mathcal{T}(z)$ onto $\Bg(\bu)|0\rangle$ is given by the same
formula as the action of $\mathcal{T}(z)$ onto $B(\bu)|0\rangle$ for arbitrary parameters $\bu$.
Then it becomes obvious that if Bethe equations \eqref{BE1} are fulfilled, then
the vector $\Bg(\bu)|0\rangle$ is proportional to the on-shell Bethe vector $B(\bu)|0\rangle$ and corresponds to the same eigenvalue
$\Lambda_0$ \eqref{act-trst}. We also would like to stress that our proof is based only on the commutation relations \eqref{RTT} and
the standard property of the vacuum vector \eqref{vac}. We did not use any specific representation of the $RTT$-algebra.

The fact that the vector $\Bg(z)\Bg(\bu)|0\rangle$ does not contribute to the action \eqref{ActtADB} also can be seen from rather general consideration.
Original operator $B(u)$ acting on $|0\rangle$ creates a state with one excitation usually called a magnon.
Action of $n$ operators $B$ gives a state with $n$ magnons. The action of the operator $\tr T(z)$ on $B(\bu)|0\rangle$ with a set $\bu$ of cardinality $\#\bu=n$ does not change the number of magnons, what can be easily seen from \eqref{ActADB}. At the same time,
the operator $\Bg(u)$ is a linear combination \eqref{Bg}. Therefore, $\Bg(\bu)|0\rangle$ is a linear combination of states with
different number of magnons. It is clear, however, that the maximal number of magnons in an individual state of this linear combination cannot exceed $n$.
It is also clear that the action of the operator $\tr T(z)$ on $\Bg(\bu)|0\rangle$ cannot change this maximal number.
On the other hand, the vector $\Bg(z)\Bg(\bu)|0\rangle$  contains a state with $n+1$ magnons. Due to the above considerations,
the action of $\tr T(z)$  cannot produce such the state. Hence, the coefficient of $\Bg(z)\Bg(\bu)|0\rangle$ must vanish, as
we have seen by the direct calculation.

This consideration stresses once more the importance of the condition $\tr \Tg(z)=\tr T(z)$.

\section{Second proof\label{S-SP}}

In this section we give one more proof of the property \eqref{tEigStat}. To do this, we need to improve our
convention on the shorthand notation. First, we introduce a rational function $h(u,v)$ as
\be{h}
h(u,v)=\frac{f(u,v)}{g(u,v)}=\frac{u-v+c}{c}.
\ee
We will consider partitions of the sets $\bu$ and $\bw=z\cup\bu$ into subsets. A notation $\bu\Rightarrow\{\bu_{\so},\bu_{\st}\}$ means
that the set $\bu$ is divided into two subsets $\bu_{\so}$ and $\bu_{\st}$ so that $\bu_{\so}\cup\bu_{\st}=\bu$ and
$\bu_{\so}\cap\bu_{\st}=\emptyset$. Similar notation will be used for other partitions. The order of the elements in each subset is not essential. We extend the convention on the
shorthand notation \eqref{shn} to the products over subsets, for example,
\be{Shn2}
a(\bu_{\so})=\prod_{u_j\in\bu_{\so}}a(u_j),\qquad  d(\bw_{\st})=\prod_{w_j\in\bw_{\st}}d(w_j),\qquad
f(\bu_{\so},\bu_{\sth})=\prod_{u_j\in\bu_{\so}}\prod_{u_k\in\bu_{\sth}}f(u_j,u_k).
\ee
By definition, any product over the empty set is equal to $1$. A double product is equal to $1$ if at least one of the sets
is empty.

To illustrate the use of this notation we give here equations \eqref{ComADB} (applied to the vacuum vector $|0\rangle$)
as sums over partitions. Let $\bw=z\cup\bu$. Then
\be{acADpar}
\begin{aligned}
&A(z)B(\bu)|0\rangle=\sum_{\bw}a(\bw_{\so})\frac{f(\bw_{\st},\bw_{\so})}{h(z,\bw_{\so})}B(\bw_{\st})|0\rangle,\\
&D(z)B(\bu)|0\rangle=\sum_{\bw}d(\bw_{\so})\frac{f(\bw_{\so},\bw_{\st})}{h(\bw_{\so},z)}B(\bw_{\st})|0\rangle.
\end{aligned}
\ee
The subscript of the sum symbol shows that the sums are taken over partitions of the set $\bw$. In \eqref{acADpar} this set is divided into subsets
$\bw\Rightarrow \{\bw_{\so},\bw_{\st}\}$ so that $\#\bw_{\so}=1$. The sum is taken over all possible partitions of this type.

It is easy to see that these equations immediately follow from \eqref{ComADB}. Indeed, if $\bw_{\so}=z$, then $\bw_{\st}=\bu$,
and using $h(z,z)=1$ we reproduce the first term in \eqref{ComADB}. If $\bw_{\so}=u_k$, where $k=1,\dots,n$, then $\bw_{\st}=z\cup\bu_k$,
and we reproduce the sums over $k$ in \eqref{ComADB}.

Similarly, one can write the action of the operator $C(z)$ onto off-shell Bethe vector $B(\bu)|0\rangle$ (see e.g. \cite{BogIK93L,BPRS12b})
\be{acC}
C(z)B(\bu)|0\rangle=\sum_{\bw}d(\bw_{\so})a(\bw_{\st})\frac{f(\bw_{\so},\bw_{\sth})}{h(\bw_{\so},z)}\frac{f(\bw_{\sth},\bw_{\st})}{h(z,\bw_{\st})}
f(\bw_{\so},\bw_{\st})B(\bw_{\sth})|0\rangle.
\ee
Here the sum is taken over partitions  $\bw\Rightarrow \{\bw_{\so},\bw_{\st},\bw_{\sth}\}$ so that $\#\bw_{\so}=\#\bw_{\st}=1$.

Now we give an explicit representation of the vector $\Bg(\bu)|0\rangle$ in terms of the ordinary off-shell Bethe vectors.
\begin{prop}\label{BgtoBV}
Let $\#\bu=n$. Then
\begin{equation}\label{BgtoB}
\Bg(\bu)|0\rangle=(\kk_{11}\kk_{12})^n\sum_{\bu} (-1)^{\#\bu_{\st}}\Big(\frac{\kk_{11}}{\kk_{12}}\Big)^{\#\bu_{\sth}}d(\bu_{\so})a(\bu_{\st}) f(\bu_{\so},\bu_{\st})f(\bu_{\so},\bu_{\sth})f(\bu_{\sth},\bu_{\st})B(\bu_{\sth})|0\rangle,
\end{equation}
where the sum is taken over all partitions $\bu \Rightarrow \{ \bu_{\so},\bu_{\st},\bu_{\sth}\}$.
\end{prop}
The proof of this proposition is quite involved, therefore, we move it to appendix~\ref{Proof}.

We have already mentioned that the vector $\Bg(\bu)|0\rangle$ is a linear combination of the ordinary off-shell Bethe vectors.
Proposition~\ref{BgtoBV} explicitly describes this linear combination. Using this explicit representation, we can easily show that
only the term with $\#\bu_{\sth}=n$ survives in the sum \eqref{BgtoB}, if the set $\bu$ satisfies Bethe equations.

First we prove an auxiliary lemma.

\begin{lemma}\label{Au-Lem}
Let $\bar x$ be a set of arbitrary complex numbers $\{x_1,\dots,x_l\}$. Then
\be{triden}
\sum_{\#\bar x_{\st}=s}f(\bar x_{\st},\bar x_{\so})=\binom{l}{s}.
\ee
Here the sum over partitions is taken under restriction $\#\bar x_{\st}=s$, where $s\in\{0,1,\dots,l\}$. We also used the
shorthand notation for the double products of the $f$-functions over the subsets $\bar x_{\so}$ and $\bar x_{\st}$.
\end{lemma}

{\sl Proof}. Clearly, the sum over partitions in \eqref{triden}
gives a rational function of $\bar x$. This rational function has no poles
in the finite complex plane,  in spite of individual terms of the sum may have singularities at $x_i=x_j$. Indeed, let,
for instance, $x_1\to x_2$. Then the pole occurs if
either $x_1\in\bar x_{\so}$ and $x_2\in\bar x_{\st}$ or  $x_1\in\bar x_{\st}$ and $x_2\in\bar x_{\so}$. Consider the first case. Then we can set
$\bar x_{\so}=x_1\cup \bar x_{\qo}$ and $\bar x_{\st}=x_2\cup \bar x_{\qt}$. The term corresponding to this partition takes the form
\be{f1}
f(x_1,x_2)\sum_{\#\bar x_{\qt}=s-1}f(\bar x_{\qt},x_1)f(x_2,\bar x_{\qo})f(\bar x_{\qt},\bar x_{\qo}),
\ee
where the sum is taken over partitions $\bar x\setminus\{x_1,x_2\}\Rightarrow\{\bar x_{\qo},\bar x_{\qt}\}$ so that $\#\bar x_{\qt}=s-1$.

In the second case we set
$\bar x_{\so}=x_2\cup \bar x_{\qo}$ and $\bar x_{\st}=x_1\cup \bar x_{\qt}$. The term corresponding to this partition takes the form
\be{f2}
f(x_2,x_1)\sum_{\#\bar x_{\qt}=s-1}f(\bar x_{\qt},x_2)f(x_1,\bar x_{\qo})f(\bar x_{\qt},\bar x_{\qo}),
\ee
where the sum is taken over the same partitions as in \eqref{f1}. Obviously, the poles at $x_1\to x_2$ in \eqref{f1} and \eqref{f2}
cancel each other.

It is  also easy to see that the sum \eqref{triden}   has a finite limit, if any $x_j\to\infty$. Hence, this function is a constant, that
does not depend on any $x_j$. Then sending all $x_j\to\infty$ (for instance, $x_j=jL$, $L\to\infty$) we make all the
$f$-functions equal to $1$. The sum becomes equal to the number of partitions of $l$ elements into two subsets with fixed number of elements in the
subset $\bar x_{\st}=s$.\qed

Let us turn back to \eqref{BgtoB}.
Consider an arbitrary partition $\bu \Rightarrow \{ \bu_{\so},\bu_{\st},\bu_{\sth}\}$. Taking a product of Bethe equations \eqref{BE1} over subset
$\bu_{\st}$ we obtain
\be{BEprod}
a(\bu_{\st})f(\bu_{\so},\bu_{\st})f(\bu_{\sth},\bu_{\st})=d(\bu_{\st})f(\bu_{\st},\bu_{\so})f(\bu_{\st},\bu_{\sth}).
\ee
Substituting the product $a(\bu_{\st})$ from this equation into \eqref{BgtoB} we find
\begin{equation}\label{BgtoBBE}
\Bg(\bu)|0\rangle=(\kk_{11}\kk_{12})^n\sum_{\bu} (-1)^{\#\bu_{\st}}\Big(\frac{\kk_{11}}{\kk_{12}}\Big)^{\#\bu_{\sth}}d(\bu_{\so})
d(\bu_{\st})f(\bu_{\st},\bu_{\so})f(\bu_{\st},\bu_{\sth})
f(\bu_{\so},\bu_{\sth})B(\bu_{\sth})|0\rangle.
\end{equation}
Let $\bu_0=\bu_{\so}\cup\bu_{\st}$. Then we recast \eqref{BgtoBBE} as follows:
\begin{equation}\label{BgtoB2}
\Bg(\bu)|0\rangle=(\kk_{11}\kk_{12})^n\sum_{\bu} \Big(\frac{\kk_{11}}{\kk_{12}}\Big)^{\#\bu_{\sth}}d(\bu_0) f(\bu_0,\bu_{\sth})B(\bu_{\sth})|0\rangle
\sum_{\bu_0} (-1)^{\#\bu_{\st}} f(\bu_{\so},\bu_{\st}).
\end{equation}
Here the sum over partitions is taken in two steps. First, we divide the set $\bu$ into subsets
$\bu \Rightarrow \{ \bu_0,\bu_{\sth}\}$. Then the subset $\bu_0$ is divided once more as $\bu_0\Rightarrow\{\bu_{\so},\bu_{\st}\}$.

It is easy to see that the sum over partitions  $\bu_0\Rightarrow\{\bu_{\so},\bu_{\st}\}$ vanishes, if $\bu_0\ne\emptyset$.
Indeed, due to lemma~\ref{Au-Lem} we have
\be{BgtoB3}
\sum_{\bu_{0}}  (-1)^{\# \bu_{\st}}f(\bu_{\st},\bu_{\so})=\sum_{s=0}^{\#\bu_0}  (-1)^{s}
\sum_{\substack{\bu_{0}\\\# \bu_{\st}=s}}f(\bu_{\st},\bu_{\so})=\sum_{s=0}^{\#\bu_0}  (-1)^{s}\binom{\#\bu_0}{s}=(1-1)^{\#\bu_0}.
\ee

Thus, a nonvanishing contribution to the sum \eqref{BgtoB2} occurs for $\bu_0=\emptyset$ only. This implies  $\bu_{\sth}=\bu$, and we arrive at
\begin{equation}\label{tB-B}
\Bg(\bu)|0\rangle=(\kk_{11})^{2n}B(\bu)|0\rangle,
\end{equation}
provided Bethe equations \eqref{BE1} are fulfilled.

\section*{Conclusion \label{Conc} }

In this paper we have studied equation \eqref{tEigStat} within the framework of the ABA. We have shown that it holds for an arbitrary
ABA-solvable model possessing the $\mathfrak{gl}_2$-invariant $R$-matrix. Furthermore, we have shown that the action of the twisted
transfer matrix $\mathcal{T}(z)$ on the vectors $\Bg(\bu)|0\rangle$ and $B(\bu)|0\rangle$ are given by the same formulas for arbitrary parameters $\bu$. Therefore, it is not surprising that both these vectors become on-shell, if the Bethe equations are fulfilled.

Note that in spite of the actions of the twisted operators $\Tg_{ij}(u)$ onto the vacuum vector are different form the ones of the original
$T_{ij}(u)$, most of the standard tools of the ABA are still available. This fact is of great importance for application of certain results
of this paper to MABA, in which the twist transformation of the monodromy matrix does not preserve its trace.
In particular, in this paper, we computed the multiple action of the $\Bg$ operator on the vacuum vector in terms of the standard
off-shell Bethe vectors.
Using exactly the same technics one can find analogous
actions  of other entries of the twisted monodromy matrix $\Tg(u)$ onto the states $\Bg(\bu)|0\rangle$ \cite{BP152}.  In their turn, these action formulas lead to new multiple action formulas \cite{BPRS12b}, in which one deals with
a product of $\Tg_{ij}(z_k)$ acting on  $\Bg(\bu)|0\rangle$. These multiple action formulas are very useful for the calculation of Bethe
vectors scalar products, form factors, and correlation functions and will be given in a forthcoming publication.

In the present paper we considered integrable models with $\mathfrak{gl}_2$-invariant $R$-matrix only. However,
most of the results of the work \cite{GroLMS17} concerns the spin chains with the symmetry of higher rank. In this case,
the authors of \cite{GroLMS17} succeeded to find an operator $B^{\text{good}}(u)$ such that, on the one  hand, it allows one to
build the SoV basis, and, on the other hand, it allows one to construct on-shell Bethe vectors in the same manner as in the case
of $\mathfrak{gl}_2$ based models. This remarkable property of $B^{\text{good}}(u)$ was checked for numerous examples, however, it was not
proved. An analytical proof of this property for the models with $\mathfrak{gl}_3$-invariant $R$-matrix will be given in our forthcoming publication.


\section*{Acknowledgements} We thank V. Pasquier, B. Vallet, I. Kostov, D. Serban and F. Levkovich-Maslyuk for discussions.
N. S. would like to thank the hospitality of the Institute de Physique Th\'{e}orique at
the CEA de Saclay where a part of this work was done. S.B. was supported by a public grant as part of the
Investissement d'avenir project, reference ANR-11-LABX-0056-LMH, LabEx LMH. N.S. was supported by the
Russian Foundation RFBR-18-01-00273a.

\appendix
\section{Proof of proposition \ref{BgtoBV} }  \label{Proof}
We use induction over $n=\#\bu$. For $n=1$, we have from \eqref{vac} and \eqref{Bg}
\be{inductbas}
\Bg(u)|0\rangle=\kk_{11}\kk_{12}\Big(\frac{\kk_{11}}{\kk_{12}}B(u)-a(u)+d(u)\Big)|0\rangle.
\ee
Thus, proposition~\ref{BgtoBV} is true for $n=1$. Suppose that it holds for some $n-1$. Then
due to \eqref{Bg} we have
\be{Mact12nn-1}
\begin{aligned}
\Bg(u_n)&\Bg(\bar u_n)|0\rangle=(\kk_{11}\kk_{12})^n\Big(\frac{\kk_{11}}{\kk_{12}}B(u_n)-A(u_n)+D(u_n)-\frac{\kk_{12}}{\kk_{11}}C(u_n)\Big)\\
& \times\sum_{\bu_n} (-1)^{\#\bu_{\st}}\Big(\frac{\kk_{11}}{\kk_{12}}\Big)^{\#\bu_{\sth}}d(\bu_{\so})a(\bu_{\st}) f(\bu_{\so},\bu_{\st})f(\bu_{\so},\bu_{\sth})f(\bu_{\sth},\bu_{\st})B(\bu_{\sth})|0\rangle
\end{aligned}
\ee
Here the subscript $\bu_n$ of the sum symbol indicates that the sum is taken over partitions of the subset $\bu_n=\bu\setminus\{u_n\}
\Rightarrow\{\bu_{\so},\bu_{\st},\bu_{\sth}\}$. Thus, we
can present this action in the form
\be{form}
\Bg(\bu)|0\rangle=(\kk_{11}\kk_{12})^n (\Lambda[B]+\Lambda[A]+\Lambda[D]+\Lambda[C]),
\ee
where $\Lambda$'s are the contributions of four operators in \eqref{Mact12nn-1}:
\be{4oper}
\begin{aligned}
\Lambda[B]&=\frac{\kk_{11}}{\kk_{12}}B(u_n)\Bg(\bar u_n)|0\rangle,\qquad &\Lambda[A]&=-A(u_n)\Bg(\bar u_n)|0\rangle,\\
\Lambda[C]&=-\frac{\kk_{12}}{\kk_{11}}C(u_n)\Bg(\bar u_n)|0\rangle,\qquad &\Lambda[D]&=D(u_n)\Bg(\bar u_n)|0\rangle.
\end{aligned}
\ee

\begin{prop}
The contributions defined by \eqref{4oper} have the form
\begin{equation}\label{LaB}
\Lambda[B]=\sum_{\bu}
\frac{\mathcal{F}_{\text{part}}(\bu_{\so},\bu_{\st},\bu_{\sth})}{f(\bu_{\so},u_n)f(u_n,\bu_{\st})},
\end{equation}
\begin{equation}\label{LaA}
\Lambda[A]=\sum_{\bu}
\frac{\mathcal{F}_{\text{part}}(\bu_{\so},\bu_{\st},\bu_{\sth})}{f(\bu_{\so},u_n)}\left(1-\frac1{f(u_n,\bu_{\st})}\right) ,
\end{equation}
\begin{equation}\label{LaD}
\Lambda[D]=\sum_{\bu} \frac{\mathcal{F}_{\text{part}}(\bu_{\so},\bu_{\st},\bu_{\sth})}
{f(u_n,\bu_{\st})}\left(1-\frac1{f(\bu_{\so},u_n)}\right) ,
\end{equation}
\begin{equation}\label{LaC}
\Lambda[C]=\sum_{\bu} \mathcal{F}_{\text{part}}(\bu_{\so},\bu_{\st},\bu_{\sth})
\left(1-\frac1{f(u_n,\bu_{\st})}\right)\left(1-\frac1{f(\bu_{\so},u_n)}\right),
\end{equation}
where
\be{Fpart}
\mathcal{F}_{\text{part}}(\bu_{\so},\bu_{\st},\bu_{\sth})=(-1)^{\#\bu_{\st}}\Big(\frac{\kk_{11}}{\kk_{12}}\Big)^{\#\bu_{\sth}}
d(\bu_{\so})a(\bu_{\st})B(\bu_{\sth})|0\rangle f(\bu_{\so},\bu_{\st})f(\bu_{\so},\bu_{\sth})f(\bu_{\sth},\bu_{\st}).
\ee
The sums in \eqref{LaB}--\eqref{LaC} are taken over all possible partitions $\bu\Rightarrow\{\bu_{\so},\bu_{\st},\bu_{\sth}\}$.
\end{prop}

Observe that taking the sum of all contributions \eqref{LaB}--\eqref{LaC} we immediately arrive at proposition~\ref{BgtoBV}:
\begin{equation}\label{LaABCD}
\Lambda[B]+\Lambda[A]+\Lambda[D]+\Lambda[C]=\sum_{\bu} \mathcal{F}_{\text{part}}(\bu_{\so},\bu_{\st},\bu_{\sth}).
\end{equation}
Thus, we  should prove equations \eqref{LaB}--\eqref{LaC}.

{\sl Proof}. We begin with the simplest contribution $\Lambda[B]$.  Obviously,
\begin{equation}\label{La0}
\Lambda[B]=\sum_{\bu_n} (-1)^{\#\bu_{\st}}\Big(\frac{\kk_{11}}{\kk_{12}}\Big)^{\#\bu_{\sth}+1}
d(\bu_{\so})a(\bu_{\st})B(u_n)B(\bu_{\sth})|0\rangle f(\bu_{\so},\bu_{\st})f(\bu_{\so},\bu_{\sth})f(\bu_{\sth},\bu_{\st}).
\end{equation}
Setting here $\bu_{\qth}=u_n\cup\bu_{\sth}$ we obtain
\begin{equation}\label{La01}
\Lambda[B]=\sum_{\bu} (-1)^{\#\bu_{\st}}\Big(\frac{\kk_{11}}{\kk_{12}}\Big)^{\#\bu_{\qth}}
d(\bu_{\so})a(\bu_{\st})B(\bu_{\qth})|0\rangle \frac{f(\bu_{\so},\bu_{\st})f(\bu_{\so},\bu_{\qth})f(\bu_{\qth},\bu_{\st})}
{f(\bu_{\so},u_n)f(u_n,\bu_{\st})}.
\end{equation}
Here in distinction of \eqref{La0} we have the sum over partitions of the complete set $\bu$. However, the terms of the sum vanish as soon as $u_n\in\{\bu_{\so}\cup\bu_{\st}\}$. This is due to the fact that $1/f(\bu_{\so},u_n)f(u_n,\bu_{\st})=0$, if $u_n\in\{\bu_{\so}\cup\bu_{\st}\}$.

Equation \eqref{La01}   coincides with \eqref{LaB} up to the labels of the subsets. 

%
%
Consider now the contribution of $\Lambda[A]$. Using the first equation \eqref{acADpar} we obtain
\begin{multline}\label{La2}
\Lambda[A]=
\sum_{\bu_n} (-1)^{\#\bu_{\st}+1}\Big(\frac{\kk_{11}}{\kk_{12}}\Big)^{\#\bu_{\sth}}d(\bu_{\so})a(\bu_{\st})
f(\bu_{\so},\bu_{\st})f(\bu_{\so},\bu_{\sth})f(\bu_{\sth},\bu_{\st})\\
\times\sum_{u_n\cup\bu_{\sth}}
a(\bu_{\qo})\frac{f(\bu_{\qt},\bu_{\qo})}{h(u_n,\bu_{\qo})}B(\bu_{\qt})|0\rangle.
\end{multline}
Here the sum over partitions is organized in two steps. First, we have standard partitions $\bu_n\Rightarrow\{\bu_{\so},\bu_{\st},\bu_{\sth}\}$.
Then we combine the element $u_n$ with the subset $\bu_{\sth}$ and take the  sum over partitions $\{u_n\cup\bu_{\sth}\}\Rightarrow
\{\bu_{\qo},\bu_{\qt}\}$ so that $\#\bu_{\qo}=1$.  Substituting $\bu_{\sth}=\{\bu_{\qo}\cup\bu_{\qt}\}\setminus u_n$ in \eqref{La2} we obtain
\begin{multline}\label{La1-2}
\Lambda[A]=
\sum_{\bu} (-1)^{\#\bu_{\st}+1}\Big(\frac{\kk_{11}}{\kk_{12}}\Big)^{\#\bu_{\sth}}d(\bu_{\so})a(\bu_{\st})
\; \frac{f(\bu_{\so},\bu_{\st})f(\bu_{\so},\bu_{\qo})f(\bu_{\so},\bu_{\qt})f(\bu_{\qo},\bu_{\st})f(\bu_{\qt},\bu_{\st})}
{f(\bu_{\so},u_n)f(u_n,\bu_{\st})}\\
\times
\frac{f(\bu_{\qt},\bu_{\qo})}{h(u_n,\bu_{\qo})}B(\bu_{\qt})a(\bu_{\qo})|0\rangle,
\end{multline}
where the sum now is taken over partitions of the complete set $\bu\Rightarrow\{\bu_{\so},\bu_{\st},\bu_{\qo},\bu_{\qt}\}$
so that $\#\bu_{\qo}=1$. Note that in the sum \eqref{La2}, we had $u_n\notin\{\bu_{\so}\cup\bu_{\st}\}$. In the sum
\eqref{La1-2} this restriction formally is absent, however the terms of the sum vanish as soon as $u_n\in\{\bu_{\so}\cup\bu_{\st}\}$.
This is due to the fact that $1/f(\bu_{\so},u_n)f(u_n,\bu_{\st})=0$, if $u_n\in\{\bu_{\so}\cup\bu_{\st}\}$.

Setting here
$\{\bu_{\qo}\cup\bu_{\st}\}=\bu_{0}$  we arrive at
\begin{multline}\label{La1-3}
\Lambda[A]=\lim_{z\to u_n}
\sum_{\bu} (-1)^{\#\bu_{0}}\Big(\frac{\kk_{11}}{\kk_{12}}\Big)^{\#\bu_{\sth}}d(\bu_{\so})a(\bu_{0})B(\bu_{\qt})|0\rangle\;\frac{ f(\bu_{\so},\bu_{0})f(\bu_{\so},\bu_{\qt})f(\bu_{\qt},\bu_{0})}{f(\bu_{\so},z)f(z,\bu_{0})}
\\
\times\sum_{\bu_0}f(\bu_{\qo},\bu_{\st})g(z,\bu_{\qo}).
\end{multline}
Here we first have the sum over partitions $\bu\Rightarrow\{\bu_{\so},\bu_{0},\bu_{\sth}\}$, and then the subset $\bu_{0}$ is divided once more
as $\bu_{0}\Rightarrow\{\bu_{\qo},\bu_{\st}\}$ so that $\#\bu_{\qo}=1$. Note, that we have replaced $u_n$ by $z$ and consider the limit $z\to u_n$.
This is because the  sum over partitions of the set $\bu_0$ becomes singular, if $u_n\in\bu_0$. Of course, this singularity eventually is compensated
by the product $1/f(u_n,\bu_{0})$, however, we should replace $u_n$ by $z$ in the intermediate formula \eqref{La1-3}.

The sum over partitions of the subset $\bu_{0}$ is a partial fraction decomposition
\be{PDF1}
f(z,\bu_{0})-1=\sum_{\bu_0}f(\bu_{\qo},\bu_{\st})g(z,\bu_{\qo}).
\ee
Substituting this into \eqref{La1-3} and setting $z=u_n$ we find
\begin{multline}\label{La1-5}
\Lambda[A]=
\sum_{\bu}  (-1)^{\#\bu_{0}}\Big(\frac{\kk_{11}}{\kk_{12}}\Big)^{\#\bu_{\qt}}d(\bu_{\so})a(\bu_{0})B(\bu_{\qt})|0\rangle\\
\times\frac{ f(\bu_{\so},\bu_{0})f(\bu_{\so},\bu_{\qt})f(\bu_{\qt},\bu_{0})}{f(\bu_{\so},u_{n})}
\left(1 -\frac1{f(u_n,\bu_{0})} \right).
\end{multline}
This equation coincides with \eqref{LaA} up to the labels of the subsets.

Calculation of the contribution $\Lambda[D]$  can be done exactly in the same manner via the second equation \eqref{acADpar}. Therefore, we
omit the details and pass to the calculating the contribution $\Lambda[C]$. Using \eqref{acC} we obtain
\begin{multline}\label{La3-1}
\Lambda[C]=
\sum_{\bu_n} (-1)^{\#\bu_{\st}+1}\Big(\frac{\kk_{11}}{\kk_{12}}\Big)^{\#\bu_{\sth}-1}d(\bu_{\so})a(\bu_{\st})
\; f(\bu_{\so},\bu_{\st})f(\bu_{\so},\bu_{\sth})f(\bu_{\sth},\bu_{\st})\\
\times\sum_{\{u_n\cup\bu_{\sth}\}}
\frac{f(\bu_{\qo},\bu_{\qt})f(\bu_{\qo},\bu_{\qth})f(\bu_{\qth},\bu_{\qt})}{h(\bu_{\qo},u_n)h(u_n,\bu_{\qt})}
d(\bu_{\qo})a(\bu_{\qt})B(\bu_{\qth})|0\rangle.
\end{multline}
Here we again deal with the sum over partitions  in two steps. First, we have the partitions $\bu_n\Rightarrow\{\bu_{\so},\bu_{\st},\bu_{\sth}\}$.
Then we combine the element $u_n$ with the subset $\bu_{\sth}$ and take the  sum over partitions $\{u_n\cup\bu_{\sth}\}\Rightarrow
\{\bu_{\qo},\bu_{\qt},\bu_{\qth}\}$ so that $\#\bu_{\qo}=\#\bu_{\qt}=1$.  Substituting
$\bu_{\sth}=\{\bu_{\qo}\cup\bu_{\qt}\cup\bu_{\qth}\}\setminus u_n$ in \eqref{La3-1} we obtain
a sum over partitions of the complete set $\bu$:
\begin{multline}\label{La3-2}
\Lambda[C]=
\sum_{\bu}(-1)^{\#\bu_{\st}+1}\Big(\frac{\kk_{11}}{\kk_{12}}\Big)^{\#\bu_{\qth}}d(\bu_{\so})d(\bu_{\qo})a(\bu_{\st})a(\bu_{\qt})
B(\bu_{\qth})|0\rangle\\
\; \frac{f(\bu_{\so},\bu_{\st})f(\bu_{\so},\bu_{\qo})f(\bu_{\so},\bu_{\qt})
f(\bu_{\so},\bu_{\qth}) f(\bu_{\qo},\bu_{\st})f(\bu_{\qt},\bu_{\st})f(\bu_{\qth},\bu_{\st})f(\bu_{\qo},\bu_{\qt}) f(\bu_{\qo},\bu_{\qth})f(\bu_{\qth},\bu_{\qt})}
{f(\bu_{\so},u_n)f(u_n,\bu_{\st})h(\bu_{\qo},u_n)h(u_n,\bu_{\qt})}.
\end{multline}
Here the sum is taken over partitions $\bu\Rightarrow\{\bu_{\so},\bu_{\st},\bu_{\qo},\bu_{\qt},\bu_{\qth}\}$ so that $\#\bu_{\qo}=\#\bu_{\qt}=1$.
The restriction $u_n\notin\{\bu_{\so}\cup\bu_{\st}\}$ automatically holds, as it was in the case of the contribution $\Lambda[A]$.
Setting $\{\bu_{\so}\cup\bu_{\qo}\}=\bu_{0}$ and $\{\bu_{\st}\cup\bu_{\qt}\}=\bu_{0'}$ we find
\begin{multline}\label{La3-3}
\Lambda[C]=\lim_{z\to u_n}
\sum_{\bu}(-1)^{\bu_{0'}}\Big(\frac{\kk_{11}}{\kk_{12}}\Big)^{\#\bu_{\qth}}d(\bu_{0})a(\bu_{0'})B(\bu_{\qth})|0\rangle
\; \frac{f(\bu_{0},\bu_{0'})f(\bu_{0},\bu_{\qth})f(\bu_{\qth},\bu_{0'})}
{f(\bu_{0},z)f(z,\bu_{0'})}\\
\times\sum_{\bu_{0}} f(\bu_{\so},\bu_{\qo})g(\bu_{\qo},z)  \sum_{\bu_{0'}} f(\bu_{\qt},\bu_{\st}) g(z,\bu_{\qt}).
\end{multline}
Here we first have the sum over partitions $\bu\Rightarrow\{\bu_{0},\bu_{0'},\bu_{\sth}\}$, and then the subsets $\bu_{0}$ and $\bu_{0'}$
are respectively divided once more
as $\bu_{0}\Rightarrow\{\bu_{\qo},\bu_{\so}\}$ and $\bu_{0'}\Rightarrow\{\bu_{\qt},\bu_{\st}\}$ so that $\#\bu_{\qo}=\#\bu_{\qt}=1$.
We also replaced $u_n$ by $z$ for the same reasons as in \eqref{La1-3}.

The sum over partitions of the subset $\bu_{0'}$ was already considered (see \eqref{PDF1}), the sum over partitions of the subset $\bu_{0}$
is the following partial fraction decomposition
\be{subsum2}
\sum_{\bu_{0}} f(\bu_{\so},\bu_{\qo})g(\bu_{\qo},z)=f(\bu_{0},z)-1.
\ee
Substituting this into \eqref{La3-3} and setting $z=u_n$
we arrive at
\begin{multline}\label{La3-4}
\Lambda[C]=
\sum_{\bu}(-1)^{\bu_{0'}}\Big(\frac{\kk_{11}}{\kk_{12}}\Big)^{\#\bu_{\qth}}d(\bu_{0})a(\bu_{0'})B(\bu_{\qth})|0\rangle
\; f(\bu_{0},\bu_{0'})f(\bu_{0},\bu_{\qth})f(\bu_{\qth},\bu_{0'})\\
\times \left(1- \frac1{f(\bu_{0},u_n)} \right)
\left(1-\frac1{f(u_n,\bu_{0'})}  \right),
\end{multline}
what coincides with \eqref{LaC} up to the labels of the subsets.

Thus, all the four actions \eqref{LaB}--\eqref{LaC} are proved, and taking the sum of these equations we obtain
the statement of proposition~\ref{BgtoBV} for $\#\bu=n$. This completes the inductive step.

\end{document}